# Motorcycle System for Optimum Road Safety with Anti-theft Capability

Carlo H. Godoy Jr.
College of Industrial Technology,
Technological University of the Philippines
Manila, Philippines

**Abstract:-** Due to road traffic accidents, 6941 Filipinos died in 2010, and thousands more were wounded or disabled. Head and neck injuries are the main cause of death, severe injury, and motorcycle users ' disabilities. Motorcycle users make up a large proportion of those on the road who were killed. The study's main purpose is to develop an MCU Based Motorcycle System for Optimum Road Safety with Anti-theft Capability that will help motorcycle riders to be safe while travelling in national roads. The researchers will be using the prototyping methodology where in a prototype is built according to the initial requirements gathered from the motorists themselves. The expected result of the proposed methodology is the system will be utilizing the different function of each modules to ensure that the riders will be able to detect and avoid possible danger while on the road. As a result of different literature in relation to each module, the system is expected to provide a new leap to ensure the safety of all riders here in the Philippines. Future studies will ensure the development of the system, provide testing and improve the system's functionality depending on the test result. Due to the high increase in the number of cars and motorcycle travelling on national road, the percentage of accidents also is getting higher. In line with that, the proposed system is expected to lessen the percentage of accident by avoiding the possible cause of it.

## I. INTRODUCTION

Metropolitan areas grew and evolved in just a few decades and are becoming, rapidly dominated by cars, motorcycle and many other automobile across the world both in developed and undeveloped nations [1]. A good transport mix typically occurs in cities of developed nations, that is, the existence of non-motorized and private motor vehicles and a good range of public transport systems, including buses of various sizes as well as train and monorail options. Motorcycles are part of the critical mode of transport in most developing countries. In the case of the Philippines, there are tricycles as well as "habal-habal" or "motorcycle taxi" in the form of local public transport. Filipinos are known to be risk-takers –choices in life and practice in their job pursuits. In their choice of transportation, it is no different. Nevertheless, risks in this area have consequences not only for one's personal health and safety, but also for the wider public.

The motorcycle industry in the Philippines is still in its infancy stage with regard to the markets of neighboring countries in the South East Asia region [2]. However, as more Filipinos begin to appreciate the convenience and practicality of owning a motorcycle, unit sales continue to grow substantially. The Participants Association Motorcycle Development Program Inc. (MDPPA) recorded an annual sales growth of 16 percent in 2017, reaching total annual sales of 1,319,084 motorcycle units.

Over the past several years, it has been an upward trend and annual sales are projected to reach 2,000,000 units by 2020. While there is an increase in the population of motorcyclists, it is quite disappointing that the general public still views motorcycles negatively [3]. It also does not help that news programs on television sometimes show motorcycles in a bad light. There are very few, if any, studies on how motorcycles in a growing economy play an important role. Some news reports, on the contrary, say they are unsafe and vulnerable to accidents, talking only about those involved in road mishaps. Nonetheless, current government reports on road accidents tend to be showing otherwise.

The Metro Manila Accident Reporting and Analysis System (MMARAS) is a program created by the Metropolitan Manila Development Authority (MMDA)-Traffic Discipline Office-Traffic Engineering Center (MMDA-TDO-TEC) Road Safety Unit (RSU) in 2005 in collaboration with and assistance from the Philippine National Police Department of Police Traffic Investigation (PNP). The purpose is to maintain a Metro Manila database of recorded road accidents [2]. The number of motorcycles involved in a road crash is 24,058 in the MMARAS 2017 survey, whereas the number of cars is 110,653. This excludes Jeeps, taxis, vans, trucks and buses from the public utility. All non-motorcycle vehicles involved in road accidents are 175,206 when counted.

The Department of Transportation (DOTr) recently established a Technical Working Group (TWG) to assess the feasibility of a motorcycle-for-hire or motorcycle taxi as a public transportation alternative (Department of Transportation, n.d.). There is an inherent weakness as a mode of transportation in a motorcycle as demonstrated by proven safety risk proof. While the current data is not directly applicable to motorcycle taxis, it is fairly accurate to quantify safety risks for all motorcycle users in the Philippines. Motorcyclists, in addition, are among the most vulnerable road users (VRUs), along with pedestrians and cyclists because they constitute "more than half (54%) of all road traffic deaths," according to the World Health Organization (WHO)'s 2018 Global Status Report for Road Safety.

In 2013, 4,668 motorcyclists were killed in traffic accidents in motor vehicles— a 6 percent drop from the 4,986 motorcyclists killed in 2012. An estimated 88,000 motorcyclists were injured in 2013, a 5% decrease from 93,000 injured motorcyclists in 2012. During 2013, during fatal crashes, two-wheeled motorcycles made up 93 percent of all motorcycles. Motorcyclists accounted for 14 percent of all traffic fatalities in 2013, 4 percent of all injured people, 18 percent of all (driver and passenger) casualties, and 4 percent of all injured occupants. Of the 4,668 motorcyclists killed as a result for traffic collisions, 94% (4,399) were riders and 6% (269) were passengers. In their latest report for 2020, roughly 1.35 million people die yearly from road traffic crashes (World Health Organization, 2020).

In line with the data from the World Health Organization, it is indeed a worldwide concern to boost the safety of the motorist since the government doesn't have the capacity to take the freedom of motorist not to use a motorcycle as a mode of transportation. The health of motorcycle riders demands that the entire road transport system is sufficiently well-oiled to work seamlessly to ensure the highest safety standards. Hence the proponents will create a system that will help boost the safety of the motorist here in the Philippines.

## II. BACKGROUND OF THE PROBLEM

According to Inquirer.net (2012), as a result of road traffic accidents, 6941 Filipinos died in 2010, and thousands more were wounded or disabled. Head and neck injuries are the main cause of death, severe injury, and motorcycle users' disabilities. Motorcycle users make up a large proportion of those on the road who were killed. Metro Manila data indicate that motorcycles are involved in 34% of all fatal road crashes and 37% of non-fatal road traffic collisions [7]. 56 percent of motorcycle accident deaths are caused by crashes involving motorcycles and other vehicles. The car hits the vehicle from the front—78 percent of the time in the vast majority of these collisions. Only 5% of the time the car strikes the motorcycle from the rear. Head-on accidents between the car and the motorcycle are often fatal to the motorcyclist [8].

To motorcyclists, the single most dangerous situation occurs when cars turn left. Such injuries constitute 42 percent of all motorcycle and car accidents. Normally, when the motorcycle is: 1.) going straight through an intersection; 2.) going through the road, or; 3.) trying to overtake the vehicle. Such types of accidents are also common among two vehicles, but the smaller size of the motorcycle makes the turning vehicle even less noticeable. Motorcycles passing cars on the same lane are even more vulnerable —cars are not anticipating these motorcycle maneuvers and are often shocked by them. Most often, the crash will result in a car that strikes another vehicle when making a left turn. Nevertheless, the motorcyclist may be partially responsible for the accident if the motorcyclist was speeding or in the wrong lane [8].

Lane splitting happens when a motorcycle moves, usually in traffic jams, between two lanes of stationary or slowly moving vehicles. Lane splitting, due to several factors, is a common cause of motorcycle accidents: (1) the cars ' close proximity to the motorcycle; (2) the reduced space that the motorcycle has to maneuver; and (3) the fact that the cars do not expect that any vehicle or motorcycle can travel through them in a slow or interrupted manner. About half of a single motorcycle accident is caused by the use of speed or alcohol. This figure is not shocking, and these causes also play a major role in automobile and other vehicle accidents. Because the driver does not get much safety from bikes, however, crashes involving speed or alcohol are much more likely to lead to death or serious injury. Motorcycles colliding with fixed objects constitute 25% of motorcyclist deaths, but only 18% of car crash deaths. Again, because the motorcyclist is not surrounded by a metal box and is likely to be thrown far and hard, when riding a motorcycle, such accidents are more deadly [8].

Regarding road hazards, motorcycles face higher risks than cars and other vehicles. Because of the motorcycle's smaller size and less stable existence, potholes, dead animals, slippery pavement surfaces, uneven lane heights, and other irregularities or unpredictable road items pose a serious safety hazard to motorcycles [8].

## III. OBJECTIVE, SCOPE AND LIMITATION OF THE STUDY

*A. Objective*
　　The general objective of the study is to develop an *MCU Based Motorcycle System for Optimum Road Safety with Anti-theft Capability.* for motorcycle riders in the Philippines to avoid road accidents.Specifically, it aims to: (1) Develop a module that will help motorcycle riders avoid motorcycles in head-on collisions; (2) Develop a module that will help motorcycle riders avoid cars making left-hand turns and motorcycle lane splitting; (3) Develop a module that will help motorcycle riders avoid accidents cause by over speeding and driving while in the influence of alcohol; (4) Develop a module that will help motorcycle riders avoid accidents cause by road hazards like human and animals suddenly crossing the road lane; (5) Develop a module that will help motorcycle riders avoid accidents caused by gas leak; (6) Develop a module that will help motorcycle riders get help during minor accidents; (7) Develop a module that will help motorcycle riders avoid accident cause by overtaking; (8) Develop a module that will help motorcycle riders monitor if the motorcycle is moving from one place to another without their knowledge; (9)Test and improve the quality of the system that will be based solely on its functionality, reliability and compatibility; and (10) Evaluate the performance of the system using the criteria of ISO 25010.

*B. Scope and Limitations*

The **Motorcycle System for Optimum Road Safety with Anti-theft Capability** will help the motorcycle riders to have a motorcycle safety system that. The system will be able to avoid road accidents that is being faced by motorist every day. The proposed system can: (1) help motorcycle riders avoid motorcycles in head-on collisions; (2) help motorcycle riders avoid cars making left-hand turns; (3) help motorcycle riders avoid accidents cause by over speeding and driving while in the influence of alcohol; (4) help motorcycle riders avoid accidents cause by road hazards like human and animals suddenly crossing the road lane; (5) help motorcycle riders avoid accidents caused by gas leak; and (6)can help motorcycle riders get help during minor accidents by notifying local police stations about what happened.

On the other hand, the proposed system has a couple of limitation. The proposed system is: (1) only specific for motorcycle it can't be used for cars and other vehicles; (2) only an aid to avoid and prevent specific accidents programmed on the system; and (3) not capable of curing a motorist that is already on a road accident.

## IV. RELATED WORK

In this part, the researchers utilized numerous sources relevant to the research of an MCU Based Motorcycle System for Optimum Road Safety with Anti-theft Capability that includes various types of books, journal articles, scientific papers and other academic papers to further understand the etymology of the processes as well as the skills involved in discussing the study of the project. This research will serve as a guide for the developers to start accomplishing the goals needed to develop the prototype.

Filipinos are considered to be risk-takers –choices in life and action in their work pursuits. In their choice of transportation, it is no different. However, risks in this area have implications not only for one's personal health and safety, but also for the wider public. According to [8], Motorcycle accidents are more likely to result in serious injury or death, although not generally more common than other types of accidents. There were 35 times more fatalities from motorcycle accidents than from car accidents, according to the federal government, per mile traveled in 2006. In road hazards, motorcycles face higher dangers than cars and other vehicles. Because of the motorcycle's smaller size and less stable nature, potholes, dead animals, slick pavement conditions, uneven lane heights, and other irregularities or unexpected road objects pose a serious safety threat to motorcycles. The reason why the proponents is focusing on an MCU Based Motorcycle System for Optimum Road Safety with Anti-theft Capability is that road accidents is not just prone in the US government but also here in the Philippines.

Since most of the country nowadays has their own way of dealing with this, it's now time for the Philippines to have our own way of protecting our citizens. The proposed system will have a Mini PIR - HC-SR505. The passive infrared motion sensor (PIR) allows a system to detect movement of a person or animal within a 6 meter range. The sensor is based on infrared radiation returned from people and animals. In the proposed system, vehicle detection will be used to focus on motorcycle safety. As discussed on the study the system will be using a magnetometer. A magnetometer works by measuring the change in the ambient magnetic field by using a passive sensing technology to detect large ferrous objects (for example, a truck, automobile, or rail car). The sensor detects those changes when a vehicle alters that magnetic field. The range of the magnetometer will depend on the target, as with other sensors. The ability to detect vehicles reliably provides significant benefits for asset management, resource allocation, site security, and traffic control. It can be challenging to identify the right technology for your vehicle detection application, and many factors need to be considered, including task, target size, sensor range, mounting of sensors and whether the application is primarily indoor or outdoor [9].

According to Wei et al., (2018), In order to protect people's lives and prevent property damage, collision avoidance systems are critical. Here we present an industrial collision avoidance sensor system in real time, designed to avoid obstacles or people and protect highly valued equipment. A scanning LiDAR and a single RGB camera are used by the system. The LiDAR is an active sensor that can function without regard to natural lighting. It can find objects reliably by their 3D reflections. The LiDAR is monochromatic, however, and can not differentiate color-based artifacts. In addition, the LiDAR can only have one to two beams intersecting the object for objects that are far away, making reliable detection problematic. The proposed system will be using a LIDAR as well. Since it is an active sensor that can function without regard to natural lighting, the proponents will maximize this and will put it at the back of the motorcycle since the headlights of other motorcycles are primary in front.

According to Gil et al., (2018), The quality of the remote sensing approaches that allow the car traffic scene (e.g. Machine Vision, LIDARs, and RADARs) to be artificially perceived is the cornerstone of the current ADAS. Regrettably, not even in the high-end segment, these sensors that are common in cars are not implemented in the tilting vehicle market. And yet, due to the high number of tilting vehicles in the fleets at global level, the security systems that such sensors require could have a significant impact. Current trends in mobility are in favor of the use of tilting cars, partly due to their potential for: electrification, recyclability and enhancement of air quality. Tilting vehicles, however, are characterized by a high risk of injury to their users, which guarantees the implementation of new and more effective safety technologies. The technological gap in tilting vehicle sensing technologies therefore translates into a relevant safety gap. In contrary to ADAS, the proposed system will be using the same components but in a different manner.

The proposed system will be using LIDAR to counter Motorcycles in Head-On Collisions with another motorcycle and the tilt sensor on the other hand will be used to detect if the motorcycle encountered an accident.

The Transportation Department (DOTr) recently established a Technical Working Group (TWG) to assess the feasibility of motorcycle-for-hire or motorcycle taxi as an alternative for public transport. Different government agencies, stakeholders and advocates for road safety attended the TWG meetings to provide their experience and the correct policy recommendations for addressing the issue at hand [12]. In lined with the formation a Technical Working Group (TWG) to assess the feasibility of motorcycle-for-hire or motorcycle taxi as an alternative for public transport, the proposed system will have a goal to improve the quality of the safety system that a motorcycle taxi may have. The proposed system will also help motorcycle taxi's to avoid being stolen.

Theft of motorcycles is one of the country's most common cases of theft. A periodic increase in cases of stolen motor vehicles and motorcycles across the country has been registered by the Philippine National Police, there are more cases of motorcycle vehicle theft compared to car theft incidents that can be easily stolen when parked unattended. It can be used to improve its security by providing an authentication before starting the motorcycle, but there are still times where it is still being stolen. The most common way a motorcycle can be stealed is to pick it off the ground and load it into a van (Aurelio as cited by Austria et al. 2017). Through this approach, the thieves can steal the motorcycle with less chance of being caught quickly and quietly (Siler as cited by Austria et al. 2017). The proponents added a way on how to be able to give not 100% but a little help for motorist in terms of motorcycle theft. An Adafruit Ultimate GPS Breakout - 66 channel w/10 Hz updates - Version 3 will be used to help the rider track if the motorcycle is moving without their knowledge thus avoiding the possibility for the motorcycle to be stolen. The gps will monitor the location of the motorcycle and will text an update every hour to the owner of the motorcycle to ensure that it stays where it is parked.

Seraj et al., [14] stated that the technology is fast moving and spreading all over the world, even in the most remote places where people are still facing basic road transport problems. The roads will become obsolete with teleportation development, so people will have to travel quickly and safely on them until then. In the meantime, the global road network is estimated at 35, 433, 439 km1 and the number of vehicles per 1000 people is estimated at 35. There are numerous studies and surveys on road deficiencies and their impact on safety and economy. Since technology is fast moving and spreading all over the world, the proponents thought of a way on how the Philippines can cope up with this. The proponents believed that this kind of study is adaptable all over Asia and the world.

According to Lacatan et al., (2016), In the Philippines, DUI offenders will have a penalty of three months to 20 years in prison. If the violator is a non-professional driver, the first conviction would result in the loss of a 12-month license and the permanent revocation of a second offense. As for the professional drivers, there is a permanent suspension of the first breach. The violator will be subjected to a series of tests to determine whether or not he is drunk and if he fails these series of tests, he will be subjected to the Alcohol Breath Analyzer Test to determine his BAC, and if his BAC is higher than the prescribed limit, he will be arrested and his car impounded. Permanently losing a license is an issue for this day and age. IID can help prevent the driver from losing his license because once the driver loses his license, he loses his driving privilege.

Jørgenrud et al., [16] conducted a study which has a goal to prove the association between speeding and use of alcohol and medicinal and illegal drugs and involvement in road traffic crashes among motor vehicle drivers. In a Norwegian roadside survey conducted in collaboration with the police, drivers of cars, vans, motorcycles and mopeds were stopped during the period from April 2016 to April 2017. An oral fluid sample (mixed saliva) analyzed for alcohol and 39 illicit and medicinal drugs and metabolites was requested from the drivers. In addition, data were collected over the past 2 years on age, sex, and self-reported speeding tickets and RTCs. The study found out that a total of 5,031 participants were included in the study, and 4.9% tested positive for the use of one or more illicit or medicinal drugs or alcohol.

The proponents will using a sensor that will detect the breath of the rider if the rider is in influence of alcohol. The proponent will be using an Adafruit MiCS5524 CO, Alcohol and VOC Gas Sensor Breakout. This sensor can easily detect gas leak indoor. It is sensitive to CO (~ 1 to 1000 ppm), Ammonia (~ 1 to 500 ppm), Ethanol (~ 10 to 500 ppm), H2 (~ 1 - 1000 ppm), and Methane / Propane / Iso-Butane (~ 1,000++ ppm). If the rider is in the influence of alcohol, the advocates use a sensor to track the rider's breath. The supporter will use Adafruit MiCS5524 CO, Alcohol and VOC Gas Sensor Breakout. This sensor can easily detect the leakage of indoor gas. It is prone to CO (~1-1000 ppm), ammonia (~1-500 ppm), ethanol (~10-500 ppm), H2 (~1-1000 ppm), and methane / propane / iso-butane (~1000 + ppm). Aside from breath it will also detect a gas leak to make sure that motorcyclist before going on a trip is safe from gas leak. Based on the past histories of motorcycle riders, the most common problem which is unnoticeable during a trip is gas leakage.

According to Consunji et al., [17], Approximately 1.2 million people die each year from motor vehicle crashes worldwide in the 2004 World Report. Most of these deaths occur in low-and middle-income countries, with one of the most prone road crash victims being pedestrians and motorcyclists. In addition to deaths, there are annual road crashes of up to 50 million people worldwide injured or permanently disabled. The ASEAN region (Brunei, Cambodia, Indonesia, Laos, Malaysia, Myanmar, the Philippines, Singapore, Thailand and Vietnam), 75,000 people die and 4.7 million are injured annually in motor vehicle crashes with 12% (9,000) deaths and 10.5% (493,500) injuries in the Philippines. Approximately 2.6% of Philippine Gross Domestic Product (GDP) or USD$1.9 billion per year. Due to the fact that 2.6% of Philippine Gross Domestic Product (GDP) is going to the budget of helping those who are injured due to motorcycle crashes, the proponents decided to come up with an idea on how to lessen this percentage and the budget that will be deducted from that can be then endorse to other sector like the education sector to provide a more quality education.

In 2007, a researcher, Uy et al. (2011) completed a report on the effect of the increasing number of motorcycles in Metro Manila. The occurrence of motorcycle road crashes was identified based on severity, type of route, time, environment, type of collision, type of junction, traffic control, causes of road crash and classification of the persons involved. The study concluded that there was a rapid increase in the registration of motorcycles, resulting in a significant increase in road crashes for motorcycles. Every year, the average road crash rate per 10,000 vehicles was reported to double from 2002 to 2005. The increase in the registration of motorcycles was stated to be 57 percent annually. Quezon City was the city with the highest number of motorcycle road accidents.

According to Cayabyab (2017), Metro Manila's number of road crashes involving motorcycles increased significantly from 2012 to 2016 by 92 percent. Road Safety Journalism Fellow Zhander Cayabyab presents the numbers in line with the publication of Republic Act 10666 Implementing Rules and Regulations or the Children's Safety on Motorcycles Act. According to Yahoo News (2019), Road accident records in the Philippines all show the country's sorry state of traffic safety. For more than 10 years since 2005, road collisions have been on a rising trend in Metro Manila alone. With 63,072 occurrences in 2007, figures nearly doubled to 116,906 in 2018. Human error accounts for eight to nine in every ten incidents nationwide, according to the Philippine National Police (PNP). Authorities believe it leads to human error through lack of driver training. One of the most identified human error causing an accident is drunk driving.

In line with data provided by the Mapua Institute of Technology, due to increasing numbers of motorcycle registration which is directly proportional to the road crash accidents, there is a need for a technology that counter this kind of problem. By applying the method "Prevention is better than cure" the proponents believes that the system would be able to act as a system prevention rather than a cure. Due to the increase in the number of road crashes in Metro Manila, the proponents can easily say that the development of an MCU Based Motorcycle System for Optimum Road Safety with Anti-theft Capability is very timely and may be able to lessen the increasing numbers of road crashes as well.

The proponents incorporated all the ideas in all the research and excluded unnecessary things that are not relevant to the current situation here in the Philippines, in line with the various research that has been done to fully understand what project needs to be done and what problem to be solved. Some of the components from the different study where put together into one creating a unique idea that has a unique identifier from each of the discuss projects. Then the MCU Based Motorcycle System for Optimum Road Safety with Anti-theft Capability came to the picture.

The proponents elaborated the idea from the different study conducted by the University of the Philippines, Adamson University and Mapua Institute of Technology together with the different international journal articles about road crashes related to motorcycles. Furthermore, the different systems discussed was upgraded coming up with the idea of building an MCU Based Motorcycle System for Optimum Road Safety with Anti-theft Capability which is very much needed today specially for motorcycle taxis.

## V. PROPOSED METHODOLOGY

The proponents proposed system is microcontroller based, which means it will operate solely by relying on the activation of the sensors and the different modules it has. The Raspberry Pi will serve as the brain of the whole system which will process all the data that will be generated by the sensors. Raspberry Pi will process the data that sensor will generate to activate the actuators to function its operation. In line with the aforementioned concepts discussed, theories that was analyzed and deciphered, and all of the findings acquired from the related literature, as well the studies from different sources and lastly learnings that had been found useful from them, the conceptual model of the study needs to be developed as illustrated in Figure 1.

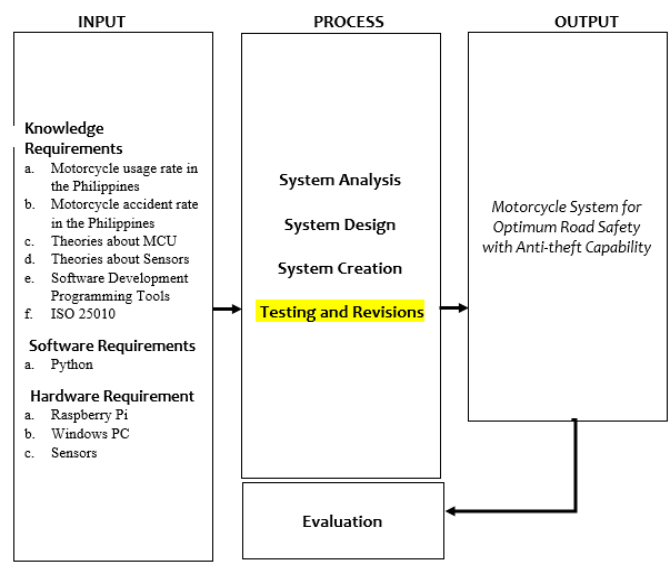

Fig 1:- Conceptual Framework of the Motorcycle System for Optimum Road Safety with Anti-theft Capability.

*A. Conceptual Framework*

Figure 1 shows the conceptual framework Integrating the learnings from the different literature review as well as studying relevant systems, the researchers came up with a newly built system that will give motorcycle riders an ease of mind while travelling in national roads.

*B. Proposed Methodology*

This part contains the methods and procedures that will be performed in the gathering of data in the proposed framework. It presents the methods of research to be used, the instruments in the gathering of data, sampling technique in evaluating of system and software development model.

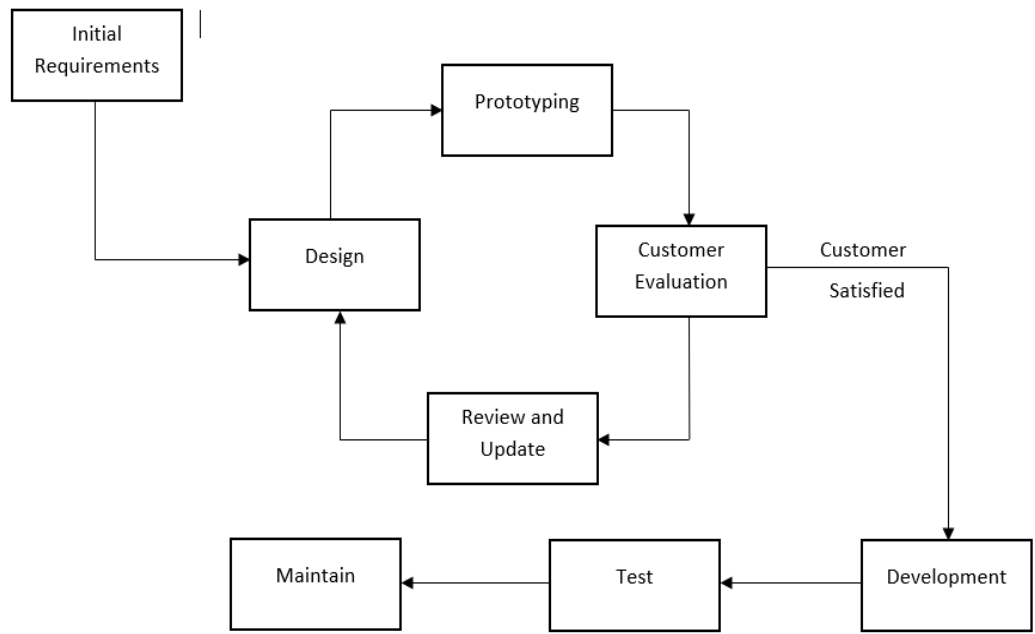

Fig 2:- Prototyping Development Methodology

As shown on Figure 2, the project will be following the prototyping methodology approach. The prototyping model is described to be a system development method use in several projects where in a prototype is built according to the initial requirements gathered from the client. It will be then tested to check if the requirements of the client has been attained. In any case depending on the result of the testing on the client's side it will be then reworked until an acceptable prototype for the client will be achieved from which the final architecture will be based on the complete system or product will be developed.

➢ *Initial Requirements*

The developer will research the topic. The researchers will also conduct a consultation with instructors that have a relative background in information technology and also in education. Also, the researchers will ask some experts that have a specialization in the field. With this, the researchers will have an initial idea of what to do on the proposed system.

➢ *Design*

In this phase, the researchers will plan and design for the whole duration of the study. The researchers will also outline the topic. The timeline or Gantt chart will be made by the researchers in this phase.

➢ *Prototyping*

In this phase of the methodology, the researchers will now organize the idea of the proposed system. With the use of the data that will be gathered by the researchers, the developer will come up with the right techniques that will be used and how to use it. In this phase, the researchers have the concrete idea of the proposed system.

➢ *Customer Evaluation*

In this part, the researchers will analyze the organized data and how it will solve the problem presented in the earlier part. The researchers will start to conceptualize the design of the proposed system. Appropriate algorithm and techniques to be used in the system will be studied. In this phase, the researchers will have the design of the system that will be used in the next phase.

➢ *Development*

In this phase, the developers will now start in creating the system. The concept made in the previous section will be applied. It includes gamification and analytics to be integrated as part of the system elements to assess the student's performance.

➤ *Testing*

　　The researchers will test the system. Riders and Computer Engineering experts will evaluate the system. The testing and evaluation phase of the system will be further discussed in the next section.

➤ *Maintain*

　　The system will be maintained depending on the data that will be provided by the riders.

C. *Project Design*

　　In developing the system, the software development model called the prototyping model will be applied as illustrated in Figure 2. The system will be divided into different modules. The different modules will be combined all together to form a complete system.

　　Once the system is done, it will be tested and evaluated. Different testing and evaluation procedures will be used. The following are the discussion of the procedures that will be done which includes the development, testing, and evaluation procedure.

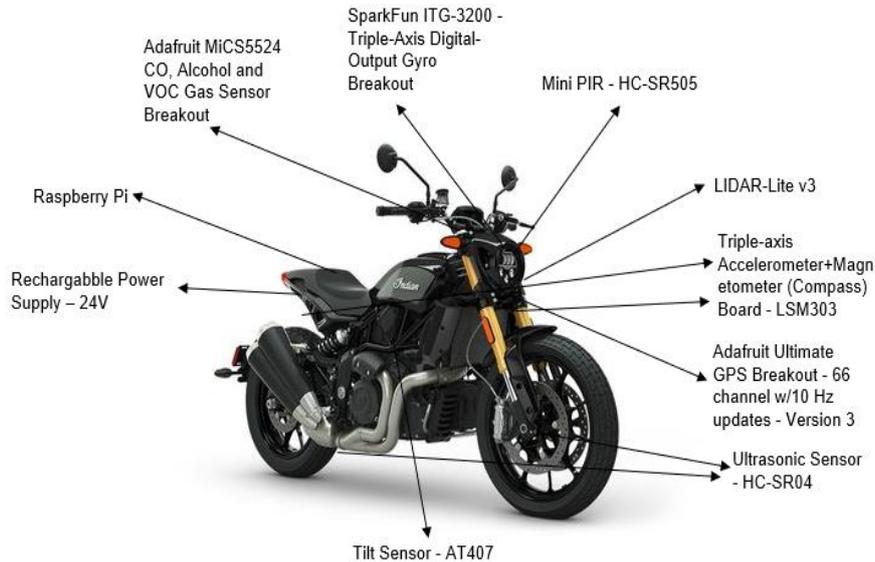

Fig 3:- Architectural Design of the MCU Based Motorcycle System for Optimum Road Safety with Anti-theft Capability

　　Figure 3 shows the proposed location of each sensors as well as the Raspberry PI. The location of each sensors is dependent on what it will detect so the Raspberry Pi can easily manage the functionality of each sensors and how it will treat the data received from it.

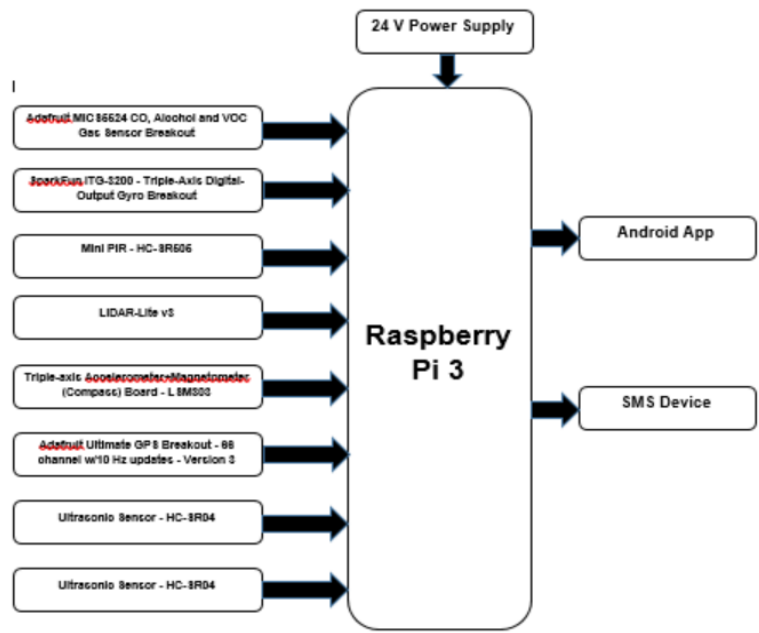

Fig 4:- Block diagram of the Motorcycle System for Optimum Road Safety with Anti-theft Capability

Figure 4 shows the relationship between each of the components. The raspberry pi will be getting its main power from the 24V power supply. The biometrics will be the one sending to the Raspberry Pi the identification of the authorized and unauthorized user. The GSM module will be the one responsible for the text notification. The WIFI module will be bridge between the microcontroller and the android application so data to be analyze will from the different sensors and module like the physiological parameter sensor, temperature, smoke, and the humidity sensor as well as the clock module. The Raspberry Pi will then send the command to the other components like the alarm module, cooling fan, servo motor, relay, solenoid lock and the door.

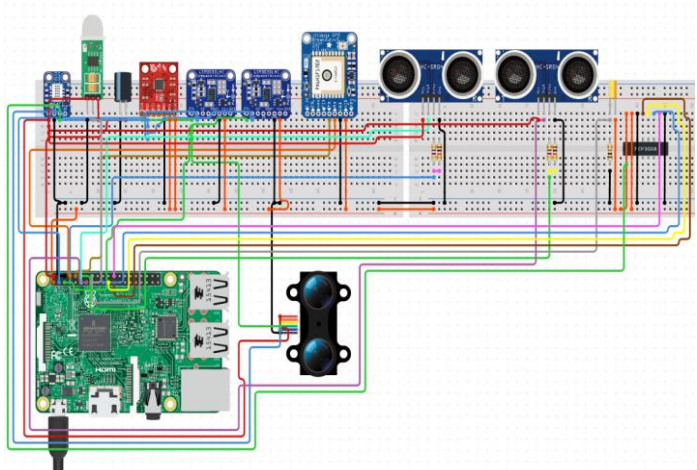

Fig 5:- Schematic Diagram of the MCU Based Motorcycle System for Optimum Road Safety with Anti-theft Capability

Figure 5 shows how the researcher will test each component and the location of the connections to be replicated in the motorcycle design shown in Figure 3.

## VI. PROPOSED OPERATION AND TESTING PROCEDURE

To ensure system quality, series of tests will be conducted for each module and by installing the system in a motorcycle to be test ridden by a rider, the system will be subjected to real-life testing.

**Proposed Functionality and Acceptability Test**. The functionality test will be used to guarantee that the application meets all the criteria and implements all the functionalities indicated in its functional requirements. Following the creation stage, the investigator will take the following measures:

- Prepared test cases for functionality for each module.
- The test cases will be executed.
- Will record the output of the test.
- The failed test cases will be analyzed and remedied.
- Failed test instances will be re-executed to check that the test instances have been remedied.

The test cases will include a set of inputs, execution preconditions and the expected results. The test case form that will be used is illustrated in Table 1.

| Test Case ID | |
|---|---|
| **Objective** | |
| **Assumptions/ Preconditions:** | |
| **Actions** | **Expected Result** |
| 1. | |
| 2. | |
| 3. | |
| **Status** | |
| **Table Head** | **Table Column Head** |
| | *Table column subhead* | *Subhead* | *Subhead* |
| copy | More table copy[a] | | |

Table 1:- Test case form

The table contains data as follows:
- Test case ID which acts as the distinctive test case identification
- Objective refers to the aim of the test case to be carried out
- Assumption/ Pre-condition identifies the original stage of the operation before it is carried out.
- Actions specify the measures to be taken in the test case
- Expected Result specifies the anticipated test outcome
- Status specifies whether or not the test case has been passed

| Module | Total No. of Test Cases |
|---|---|
| Motorcycle Riders | |
| | |
| | |
| | |
| **Total** | |

Table 2:- Over-all summary of functionality test cases

As shown in Table 2, a number of test cases will be prepared for execution to check the performance of each module. All functionality test instances will be shown in the Appendix C. The test instances will be performed during the testing stage and outcomes will be registered for each test case, classifying whether the test case is passed or failed. A condition passed implies that the real findings fulfilled the anticipated outcomes. Failed condition, on the other side, implies the case has been performed with mistakes or the real outcomes are distinct from the anticipated result.

To check the test cases, false positive and false negative, true positive true negative will be observed. This will check the Accuracy and Error of the system which will be very vital. The following formula will be used to compute the Accuracy and Error:

$$\text{Accuracy} = \frac{TP + TN}{TP + TN + FN + FP} \times 100$$

$$\text{Error} = \frac{FP + FN}{TP + TN + FN + FP} \times 100$$

If the test case fails, the error will be noted in a test incident log as shown in Table 5, indicating where applicable the error description, test case reference, severity, priority and screenshot reference. The errors in a failed test case will be fixed and the corresponding test case will be re-executed to verify that the problems have been resolved.

| No. | Description of the Error | Test Case Reference | Severity | Priority | Screenshot Reference |
|---|---|---|---|---|---|
| | | | | | |

Table 3:- Test incident log

Table 3 describes the different severity classification and priority levels that can be assigned to an error.

| Severity | Description | Priority |
|---|---|---|
| (Severity 1) | These problems are showstoppers and there is a serious obstacle to productivity. There are no workarounds and instant resolution is required. | High |
| (Severity 2) | These are issues that have impact to productivity but can be worked around | Medium |
| (Severity 3) | These are problems that affects productivity but can be addressed | Low |

Table 4:- Severity classification and priority level of errors

At the end of the proposed testing stage, the test findings will be collected and summarized. Functionality testing will be effective if 100 percent of the documented test instances with a Passed status as shown in Table 5 are performed which will be done in the next research after the system has been created.

| Test Execution (TC – Test Case) | Expected Result | Actual Result | |
|---|---|---|---|
| | | Cycle 1 | Cycle 2 |
| No of TC Executed | 100% | | |
| Results of TC | | | |
| Successful | 100% | | |
| Failed | 0% | | |
| No of TC Not Executed | 0% | | |

Table 5:- Test case execution summary

**Proposed Reliability, Suitability and Usability Test**. This test will be performed to confirm that for a particular period, the system should be able to perform the designated tasks correctly. Live testing will be performed to check the system's reliability. The system will be implemented and exposed to real-world testing in a live setting. The following measures are to be taken: (1) Live deployment of the scheme will be initiated; (2) Will request possible end-users which will include Computer Engineers, students, teachers and motorcycle riders; (3) Will ask end-users to forward back to the investigator the problems experienced; (4) Will collect the outcomes and tabulate them.

The issues that will be sent by the end-users will be noted down in a table illustrated in Table 6, which will show the following: affected module, description of the encountered issue, who reported the problem encountered and its severity and priority. Fixes will be implemented to each reported problem and end users will be asked to verify if the problem has been resolved in the future studies.

| Module Affected | Description of the Issue Encountered | Reported By | Severity | Priority |
|---|---|---|---|---|
| | | | | |
| | | | | |

Table 6:- Proposed live testing results form

The investigator will also perform the following measures in relation to live system for reliability testing: (1) Preparation of test cases for reliability test; (2) Will perform every test cases; (3) Records the output of every test cases.

Any reliability test case that will be documented using the test case form which can be seen in Table 1. The found problems will be documented in a test incident log as can be seen in Table 3 and a summary of the execution of the reliability test case will be collated using the format shown in Table 5. The total number of test cases will be executed and will be summarized in a table similar to the illustrated table in Table 7, to check whether the system is reliable under different instances. All test instances that will be performed for reliability should be discovered in the future research once the system has been created.

| Test Case ID | Objective |
|---|---|
| AR001 | |
| AR002 | |
| AR003 | |
| AR004 | |
| AR005 | |
| AR006 | |
| AR007 | |

Proposed reliability test case summary

## VII. CONCLUSION AND FUTURE WORKS

The rise in ownership of Motorcycle in the Philippines has resulted in an increase in the number of accidents associated with its use. Approximately 6.7% of trauma admissions at the Philippine General Hospital (PGH) from 2004-2006 were victims of Motorcycle accidents ( Consunji et al. as cited by Seva 2017) and 40.4% of injuries in the sample of 156 patient records were related to MC accidents in a case study conducted in 2013 at the Manila Doctors Hospital (O'Connor and Ruiz as cited by Seva 2017). Motorcycle riders are involved in various types of accidents, from minor to fatal. Although there is an increasing trend in the number of reported accidents involving MCs in the Philippines, it is assumed that there are many cases that remain especially minor unreported.

As per the Carmudi Insider, motorcycle is not just used for personal reasons but also for business ones. Since the market of motorcycle is increasing, the mindset for safety should increase as well. According to Carmudi Insider (2018) Business-use motorcycles, which reported sales growth of 18 percent at 202,098 units last year compared to 170,979 units in 2016, are expected to continue on an upward trajectory due to the increasing popularity of the cocooning trend. This trend is described as the preference of consumers to do activities such as working, entertaining and relaxing at home rather than outside, especially millennials. With more people now finding it productive to complete tasks, even shopping, from the comfort of their own homes, wherever internet connection feeds the trend even further.

For future studies, In line with data provided by the University of the Philippines and Carmudi Insider, there is a need for a technology to combat this type of problem due to the increasing number of motorcycle registration that is directly proportional to road crash accidents. This kind of technology will have an aim to lessen the percentage that was mentioned on the study conducted by the University of the Philippines. At this point, a Motorcycle System for Optimum Road Safety with Anti-theft Capability will be developed by the researcher. The developed framework and proposed methodology in this study will be used in the creation process. Once the system is developed, it must be evaluated by Computer Engineering experts using the Functionality, Reliability, Suitability and Usability of the ISO 25010. The acceptability testing of the motorcycle riders shall be conducted as well. With this, it can assess that the system can be a good tool to improve learning experiences and students' attitudes toward lessons using gamification approach.